\documentclass[aps,prl,twocolumn,groupedaddress]{revtex4-1}

\usepackage{graphicx}
\usepackage{dcolumn}
\usepackage{bm}
\usepackage{amsmath}
\usepackage{amssymb}
\usepackage{latexsym}
\usepackage{epsfig}
\usepackage{amsbsy}
\usepackage{array}
\usepackage{amssymb}
\usepackage{setspace}
\usepackage{bm}

\def\sint{\ifmmode{- \!\!\!\!\!\! \int}
    \else{\hbox{$- \!\!\!\! \int \ $}}\fi}

\begin{document}

\bibliographystyle{apsrev4-1}

\title{Electric control of inverted gap and hybridization gap in type II InAs/GaSb quantum wells}

\author{Lun-Hui Hu$^{1,2}$}
\author{Chao-Xing Liu$^3$}
\email{cxl56@psu.edu}
\author{Dong-Hui Xu$^4$}
\author{Fu-Chun Zhang$^{1,2}$}
\email{fuchun@hku.hk}
\author{Yi Zhou$^{1,2}$}
\affiliation{$^1$Department of Physics, Zhejiang University, Hangzhou, Zhejiang, 310027, China}
\affiliation{$^2$Collaborative Innovation Center of Advanced Microstructures, Nanjing 210093, China}
\affiliation{$^3$Department of Physics, The Pennsylvania State University, University Park, Pennsylvania 16802, USA}
\affiliation{$^4$Department of Physics, Hong Kong University of Science and Technology, Clear Water Bay, Hong Kong, China}

\date{\today}

\begin{abstract}
  The quantum spin Hall effect has been predicted theoretically and observed experimentally in InAs/GaSb quantum wells as a result of inverted band structures, for which electron bands in InAs layers are below heavy hole bands in GaSb layers in energy. The hybridization between electron bands and heavy hole bands leads to a hybridization gap away from k=0. A recent puzzling observation in experiments is that when the system is tuned to more inverted regime by a gate voltage (a larger inverted gap at k=0), the hybridization gap decreases. Motivated by this experiment, we explore the dependence of hybridization gap as a function of external electric fields based on the eight-band Kane model. We identify two regimes when varying electric fields: (1) both inverted and hybridization gaps increase and (2) inverted gap increases while hybridization gap decreases. Based on the effective model, we find that light-hole bands in GaSb layers play an important role in determining hybridization gap. In addition, a large external electric field can induce a strong Rashba splitting and also influence hybridization gap.
\end{abstract}


\maketitle

Two dimensional quantum spin Hall (QSH) insulator \cite{kane_prl_2005,bernevig_prl_2006,Bernevig_science_2006,Konig_science_2007} has an insulating band gap and conducting helical edge channels at the boundary, which consist of two counter-propagating one dimensional edge modes with opposite spin. The helical edge modes (HEMs) are protected by time reversal symmetry and characterized by a $Z_2$ topological invariant. The first experiment evidence of the QSH effect was identified in transport measurements of HgTe/CdTe quantum wells \cite{Konig_science_2007}, for which a $2e^2/h$ conductance was observed in a two-terminal measurement when the Fermi energy is tuned into the bulk energy gap. Recently, increasing interests were attracted to another QSH insulator, the type II InAs/GaSb quantum wells \cite{liu_prl_2008,knez_prl_2011,knez_prl_2014,du_prl_2015}, because a more robust $2e^2/h$ conductance plateau, which surprisingly persists up to 10T for an in-plane magnetic field \cite{du_prl_2015}, was found in this system. More recent experiments also revealed strong interaction effect of HEMs (helical Luttinger liquids) \cite{li_prl_2015}, presumably due to the small Fermi velocity in this system. In addition, a strong superconducting proximity effect between InAs and superconductors \cite{mourik_science_2012} provides an interesting platform for the study of topological superconductivity and Majorana zero modes in this system.

Recent experiments\cite{dulingjie2015} by Du's group on this system have revealed a puzzling effect about the dependence of band gap on external gate voltages. To illustrate this puzzling effect, we first review electronic band structure of InAs/GaSb quantum wells.

\begin{figure}[!htbp]
  \begin{minipage}[b]{0.5\textwidth}
      \includegraphics[width=3.0in]{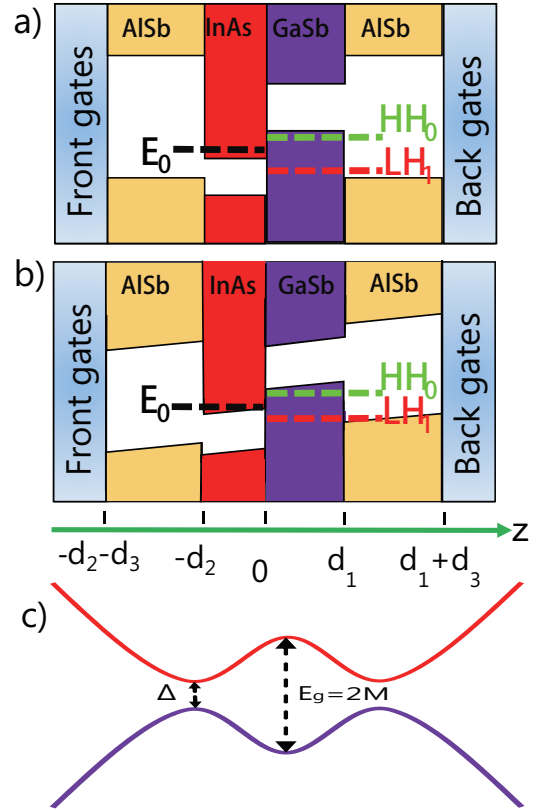}
  \end{minipage}
  \caption{\label{Fig-1-sketch-band-structure} Sketch figure of band structure for InAs/GaSb quantum well. a) Without external electric field. b) With positive external electric potential \(U\) along z-direction. c) Inverted band structure with hybridization gap (mini-gap) labelled with \(\Delta\); \(E_g=2M\) is referred to the inversion gap at \(\Gamma\) point.}
\end{figure}

The band diagram of AlSb/InAs/GaSb/AlSb QWs is shown in Fig.\ref{Fig-1-sketch-band-structure}(a), in which the conduction band minimum of InAs layer is 0.15 eV lower in energy than the valence maximum band of GaSb layer, leading to an inverted band structure. As a consequence, there is an anti-crossing between the electron sub-band in the InAs layers, denoted as \(\vert E_0\rangle\), and the heavy-hole sub-band in the GaSb layers, denoted as \(\vert HH_0\rangle\), due to hybridization between them. The anti-crossing energy dispersion is shown in Fig.\ref{Fig-1-sketch-band-structure}(c), in which the band gap at $\Gamma$ ($k=0$) is usually known as the inverted band gap, denoted as \(E_g=2M\), while the small gap at a finite momentum $k$, labeled by $\Delta$, is known as hybridization gap or mini-gap. The low energy physics for the \(\vert E_0\rangle\) and \(\vert HH_0\rangle\) sub-bands can be captured by a simple four band model \cite{Bernevig_science_2006}, which is the standard model to understand the QSH physics in this system.

The recent transport measurement \cite{dulingjie2015}  revealed the relation between the inverted gap $M$ and mini-gap $\Delta$ by independently controlling chemical potentials and asymmetric potentials in an experimental setup with front and back gate voltages, as shown in Fig.\ref{Fig-1-sketch-band-structure}(a). In their experiment, they fix the back gate voltage and tune front gate voltage continuously to reach a charge neutrality point (CNP), which corresponds to the point with equal electron and hole density ($n_0=p_0$) and can be dictated by sign change of the Hall resistance. By fitting the Hall resistance to a two-carrier model \cite{smith_book}, carrier densities and mobilities for both electrons and holes can be extracted. By tuning back gate voltage, it is shown that electron density $n_0$ at CNP can reduce from \(9\times10^{10} \text{ cm}^{-2}\) to \(5\times10^{10} \text{ cm}^{-2}\), implying that the inverted band gap $E_g$ is also reduced.
Moreover, the measurement of longitudinal conductance \(\sigma_{xx}\) in a double-gated Corbino device as a function of temperature reveals an excitation gap for different carrier density \(n_0\) at CNP. The mini-gap $\Delta$ can be extracted by fitting the temperature dependence of longitudinal resistance with the equation \(\sigma_{xx}\approx \exp\left(-\Delta/2k_BT\right)\). Surprisingly, it is found that the mini-gap \(\Delta\) increases from 0.5 meV to 3 meV (5K to 30K) as \(n_0\) decreases from \(9\times 10^{10} \text{ cm}^{-2}\) to \(5\times 10^{10} \text{ cm}^{-2}\). As the electron density \(n_0\) at CNP is further reduced below \(5\times 10^{10} \text{ cm}^{-2}\)), the mini-gap $\Delta$ is reduced.

To get more insight of  Du's experiment, we first consider the relation between electron density at CNP and mini-gap $\Delta$ in the four band model described in Ref. [\onlinecite{Bernevig_science_2006}] and [\onlinecite{liu_prl_2008}]. The CNP is determined by the momentum \(k_c=\sqrt{\frac{M}{B}}\) of the anti-crossing point between \(\vert E_0\rangle\) and \(\vert HH_0\rangle\), where $M$ is inversion gap and $B$ is the coefficient before quadratic terms. The electron density at the CNP is related to \(\mathbf{k}_c\) by \(n_0=\frac{\mathbf{k}^2_c}{2\pi}=\frac{M}{2\pi B}\), from which one can see that $n_0$ is directly proportional to inversion gap $M$. On the other hand, the mini-gap $\Delta$ is also related to ${\bf k}_c$ by $\Delta=A k_c=A\sqrt{\frac{M}{B}}$. As a result, we expect that the mini-gap should be proportional to $\sqrt{n_0}$ in the four band model, and this simple analysis is clearly in contradiction to the experimental observation described above. Thus, a study beyond the four band model is required to understand this experimental observation.

In order to explore the dependence of the inversion gap and the mini-gap on external electric fields, we perform a numerical calculation of electronic band structures with external electric fields. Our starting point is the eight-band Kane Hamiltonian $\mathcal{H}_{K}$ derived at $\Gamma$ point in the framework of $\mathbf{k}\cdot\mathbf{p}$ theory\cite{Kane1957249}, together with an external electric field term. The full Hamiltonian is given by
\begin{align}\label{eq:ham_full}
     \mathcal{H}_{\text{full}} &= \mathcal{H}_{\text{K}} + \mathcal{V}_{\text{ext}} \\
     \mathcal{V}_{\text{ext}}  &= \left\lbrack \left(\frac{U}{L}\right) \text{eV}\cdot\text{\r{A}}^{-1} \right\rbrack (z\,\text{\r{A}}).
\end{align}
The eight-band Kane Hamiltonian $\mathcal{H}_{K}$ \cite{kane_book12,winkler_book} is written in the basis $\vert\lambda\rangle (\lambda=1,2,\cdots,8)$, labelling 2 s-orbital bands and 6 p-orbital bands. $|\lambda=1\rangle$ and $|2\rangle$ are for two s-orbital bands $\vert \Gamma^6,\pm1/2\rangle$, $|3\rangle$ and $|6\rangle$ are for high hole bands $\vert\Gamma^8,\pm3/2\rangle$; $|4\rangle$ and $|5\rangle$ are for light hole bands $\vert\Gamma^8,\pm1/2\rangle$; and $|7\rangle$ and $|8\rangle$ are for $\vert\Gamma^7,\pm1/2\rangle$.
The structure parameters for InAs, GaSb and AlSb at $T=0 K$ are listed in Table I of Appendix I in the supplementary materials.  In Eq. (1), \(\mathcal{V}_{ext}\) denotes the external electric potential, $U$ is the potential drop and $L=d_1+d_2+2d_3$ is the total width of the QWs, and  \(d_1, d_2, d_3\) are the widths of GaSb, InAs and AlSb layers, respectively,  as shown in Fig.\ref{Fig-1-sketch-band-structure}. For the quantum well problem, we need to solve the following eigen-equation
\begin{align}\label{eq:eigenvalues-kane}
     \mathcal{H}_{\text{full}}(-i\partial_z,k_x,k_y)|f(z)\rangle = E |f(z)\rangle
\end{align}
with the envelope function vector $|f(z)\rangle=\left(f_1(z),f_2(z),\cdots,f_8(z)\right)^{T}$, which may be expanded  numerically in terms of the plane waves.  The detailed technique information is included in Appendex I and II.

\begin{figure}[!htbp]
  \begin{minipage}[b]{0.5\textwidth}
      \includegraphics[width=3.2in]{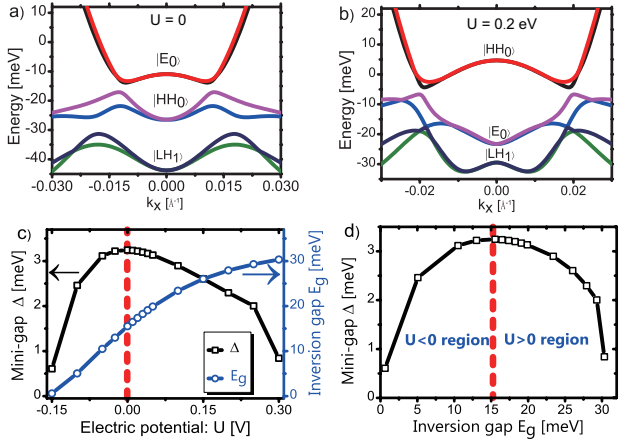}
  \end{minipage}
  \caption{\label{Fig-2-dispersion-minigap} Calculational results from eight-band Kane model.  a) Dispersion without external electric field (\(U=0\)). b) Dispersion when the external electric potential \(U=0.2\) eV. c) Mini-gap \(\Delta\) and inversion band gap \(E_g\) as functions of electric potential \(U\). d) The relationship between  \(\Delta\) and \(E_g\). In the calculations, the widths of QWs are \(d_1=d_2=\) 10 nm and \(d_3=\) 25 nm. All the other parameters can be found in Appendix I in the supplementary materials.}
\end{figure}

Our main numerical results are summarized in Fig. \ref{Fig-2-dispersion-minigap}. In Fig. \ref{Fig-2-dispersion-minigap}(a) and (b), we show two typical energy dispersions for the potentials $U=0$ eV and $U=0.2$ eV, respectively. The system is in the inverted regime for both potentials, where $|E_0\rangle$ state is below $|HH_0\rangle$ state in energy. A mini-gap $\Delta$ occurs around the momentum $k=0.0125 \text{ \r{A}}^{-1}$ for $U=0$ eV and $k=0.02 \text{ \r{A}}^{-1}$ for $U=0.2$ eV. All energy bands are spin split by Rashba spin-orbit coupling, due to the lack of inversion symmetry. With increasing the potential $U$ (positive electric field), one can see that the system is driven into more inverted regime with a larger inverted gap from 15 meV to 30 meV. Rashba spin splitting is also significantly enhanced for both conduction and valence bands. Another important feature is that with a large potential, the $|LH_1\rangle$ state also move closer to the band gap and show a strong mixing with $|E_0\rangle$ state. These features can be understood from a simple physical picture shown in Fig. \ref{Fig-1-sketch-band-structure}(b). In a positive electric field, the energies in both the heavy hole (\(\vert HH_0\rangle\)) and light hole (\(\vert LH_1\rangle\)) valence bands in GaSb layer increases, while the energy in the conduction band of InAs layer (\(\vert E_0\rangle\)) decreases. As a consequence, the inverted band gap \(E_g\) is increased, while the light hole band is pushed toward the Fermi energy.

In Fig. 2(c), we plot the inverted band gap $E_g$ and mini-gap $\Delta$ as functions of external potential $U$, which behave very differently.  $E_g$ increases monotonically as $U$ increases, while $\Delta$ has a maximum at $U=0$ and decreases as $|U|$ increases, for both positive and negative U. The relationship between  \(\Delta\) and \(E_g\) is summarized in Fig.\ref{Fig-2-dispersion-minigap}(d), from which one can see that the mini-gap \(\Delta\) decreases from 3 meV to 0.5 meV while the inverted gap $E_g$ increases from 15meV to 30meV when the external electric potential $U$ is tuned from 0eV to 0.3eV. In Du's experiment \cite{dulingjie2015}, the excitation gap, which is measured from temperature dependence of longitudinal conductance, also decreases from 3meV to 0.5meV when electron density $n_0$ at the CNP increases by around two times. Since $n_0=\frac{M}{2\pi B}$ is directly proportional to the inverted gap $E_g=2M$, we conclude that our calculations are consistent with experimental observations. We may estimate $n_0$ in our model, $n_0 = 2.15\times 10^{11}\text{ cm}^{-2}$ for electric potential $U=0\,eV$; and $n_0=6.06\times 10^{11}\text{ cm}^{-2}$ for $U=0.2\,eV$. This estimation for $n_0$ differs from the values in the experiment\cite{dulingjie2015} and this is because our quantum well configuration is not exactly the same as that in experiments. However, the qualitative change tendency for $n_0$ and $\Delta$, $E_g$ is consistent with experiments.

\begin{figure}[!htbp]
  \begin{minipage}[b]{0.5\textwidth}
      \includegraphics[width=3.3in]{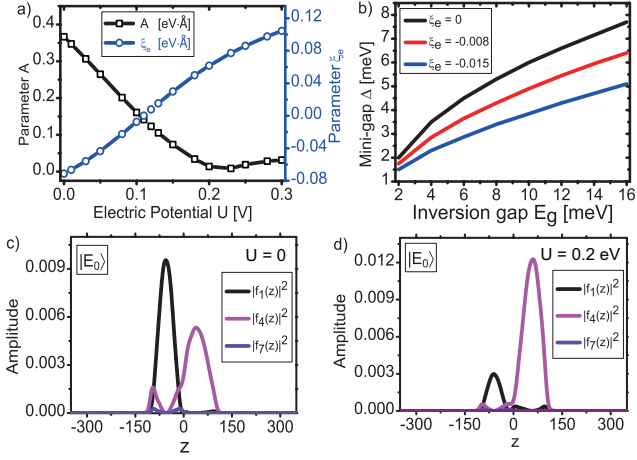}
  \end{minipage}
  \caption{\label{Fig-3-wave-func-parameter-A-xi} a) Parameters \(A\) and \(\xi_e\) in BHZ model as functions of external potential $U$, obtained by first order perturbation theory from eight band Kane model. b) Relationship between mini-gap \(\Delta\) and inversion band gap \(E_g=2M\) from BHZ model with different Rashba term \(\xi_e\).
  c) and d) show the amplitude of wave functions for \(E_0\) state at \(\Gamma\) point at \(U=0\) and \(U=0.2\) eV, respectively. Not shown are those amplitudes with zero or tiny values for other components of wave function. In the calculations, the widths of QWs are \(d_1=d_2=\) 10 nm and \(d_3=\) 25 nm. All the other parameters can be found in Appendix I in the supplementary materials.}
\end{figure}

To provide more theoretical insight to the underlying physics, we next study the four band model for InAs/GaSb QWs. A key insight is that by tuning external electric fields, not only the inverted band gap $M$ is modified, but also other parameters, such as \(A\) and \(\xi_e\), be changed. Therefore, we adopt the perturbation theory \cite{winkler_book} to derive all the parameters in the four-band model from the eight-band Kane model \cite{rothe_review_njp_2010}. We treat \(\mathcal{H}_{\text{full}}(k_z,k_x=0,k_y=0)\) including electric fields in Eq.\eqref{eq:ham_full} as non-perturbation Hamiltonian and solve the eigen energies and eigen wave functions at \(\Gamma\) point. All the other terms including non-zero \(k_x,k_y\) are regarded as the perturbation part of the Hamiltonian. Up to the first order in the momentum $k$, we find that \(A\) and \(\xi_e\) near \(\Gamma\) point can be given by
\begin{widetext}
  \begin{eqnarray}\label{eq:parameter-for-bhz-1}
     A &=& -\frac{1}{\sqrt{2}}\langle f_{E_0+,1}\vert P\vert f_{HH_0+,3}\rangle + \frac{\sqrt{3}\hbar^2}{2m_0}\langle f_{E_0+,4}\vert \{\gamma_3^\prime,k_z\} + \lbrack k_z,\kappa\rbrack \vert f_{HH_0+,3}\rangle \\
     \xi_e &=& -i\frac{1}{\sqrt{6}} \left\lbrack  -\langle f_{E_0+,4}\vert P\vert f_{E_0-,2}\rangle + \langle f_{E_0+,1}\vert P\vert f_{E_0-,5}\rangle \right\rbrack - i\frac{\hbar^2}{2m_0}\langle f_{E_0+,4}\vert \lbrack \kappa,k_z\rbrack \vert f_{E_0-,5}\rangle.
          \label{eq:parameter-for-bhz-2}
  \end{eqnarray}
\end{widetext}
where $\kappa,  P$ and $\gamma_3'$ are material dependent parameters, whose values are listed in Appendix I in the supplementary materials. Note that \(\gamma_3^\prime\) is a re-normalized parameter\cite{winkler_book}. In the above equations, the eigen wave function is expanded as $|f_{\alpha}\rangle=\sum_{\lambda}f_{\alpha,\lambda}(z)|\lambda\rangle$ and the inner product is defined as $\langle f_{\alpha,\lambda}|V(z)|f_{\alpha',\lambda'}\rangle=\int dz f^*_{\alpha,\lambda}(z) V(z) f_{\alpha',\lambda'}(z)$ for any function $V(z)$. From Eq. (\ref{eq:parameter-for-bhz-1}) and (\ref{eq:parameter-for-bhz-2}), one can directly compute parameters $A$ and $\xi_e$ as functions of the electric potential $U$. The results are summarized in Fig. \ref{Fig-3-wave-func-parameter-A-xi}(a). The parameter $A$ decreases from $0.367$ eV$\cdot\text{\r{A}}$ to $0.034$ eV$\cdot\text{\r{A}}$  as $U$ increases from $0$ eV to $0.2$ eV, then saturates as $U$ further increases. On the other hand, $\xi_e$ monotonically increases and changes its sign from negative to positive at $U \approx 0.1$ eV, which may be attributed to the opposite contribution to the Rashba coupling under the positive electric potential $U$, as compared to the Rashba coupling from intrinsic structure asymmetry of our quantum well structure.  As we shall analyse below, the decrease of the mini-gap $\Delta$ as $U$ changes from $0$ eV to $0.2$ eV is mainly due to the rapid drop of $A$, and the decrease of $\Delta$ as $U$ increases beyond $0.2$ eV is due to the rapid increase of $\xi_e$.

We first consider the region $0\,\text{eV}<U<0.2\,\text{eV}$. Since $\Delta=A\sqrt{M/B}$, while $M$ increases as $U$ increases, $A$ decreases more rapidly and plays an important role to lead the reduction of $\Delta$.  More quantitatively, we find \(\frac{A(U=0)}{A(U=0.2)}\sim 11\) and \(\frac{M(U=0)}{M(U=0.2)}\sim0.6\), leading to a change of mini-gap \(\frac{\Delta(U=0)}{\Delta(U=0.2)}\sim 8\). This  estimate explains why mini-gap $\Delta$ decreases even though the inversion gap $E_g$ increases in the region $0\,\text{eV}<U<0.2\,\text{eV}$ in Fig.\ref{Fig-2-dispersion-minigap}(c).

We next consider the region \(0.2\,\text{eV}<U<0.3\,\text{eV}\), in which the parameter $A$ remains almost unchanged hence does not play much role to the change of the minigap as we can see from Fig. \ref{Fig-3-wave-func-parameter-A-xi}(c). On the other hand,  the Rashba term \(\xi_e\) becomes more important in this region. Due to the intrinsic structure asymmetry of InAs/GaSb QWs, there is a non-zero Rashba coupling at $U=0$, which is negative in the present convention.  A positive electric potential $U$ gives rise to a Rashba coefficient of opposite sign, so that the net Rashba coefficient $\xi_e$ increases and becomes positive at \(U > 0.1\,\text{eV}\).  In Fig. \ref{Fig-3-wave-func-parameter-A-xi}(b), we plot $\Delta$ vs $E_g$ for three values of $\xi_e$.  Clearly, $\Delta$ also depends on $\xi_0$. Thus, we attribute the reduction of $\Delta$ in the region $U> 0.2\,\text{eV}$ to the increased value of $\xi_e$.

Let us return to the dependence of the parameter $A$ on the external electric potential $U$.  This can be understood from the composition of the eigen wave function $|E_0\rangle$, as shown in Fig. \ref{Fig-3-wave-func-parameter-A-xi}(c) and (d). As we can see from Fig. \ref{Fig-1-sketch-band-structure}(b), the light hole sub-band $|LH_1\rangle$ moves close to the electron sub-band $|E_0\rangle$ as $U$ increases from zero. As a consequence, there is a strong hybridization between the components $|\Gamma_6,\pm\frac{1}{2}\rangle$ and $|\Gamma_8,\pm\frac{1}{2}\rangle$.
In particular, we find at $U=0.2\,\text{eV}$, the eigen state $|E_0,+\rangle$ mainly consists of the components $|\Gamma_8,\frac{1}{2}\rangle$ ($f_{E_0+,4}$), in sharp contrast to the case of $U=0$, where $|\Gamma_6,\frac{1}{2}\rangle$ ($f_{E_0+,1}$) is the dominant component. From Eq. (\ref{eq:parameter-for-bhz-1}), both $f_{E_0,1}$ and $f_{E_0,4}$ terms may contribute to the parameter $A$. However, numerical calculations show that the contribution to $A$ from the term \(\langle f_{E+,1}\vert P\vert f_{H+,3}\rangle\)  is about ten times larger than that from the term \(\langle f_{E+,4}\vert \{\gamma_3^\prime,k_z\} + \lbrack k_z,\kappa\rbrack \vert f_{H+,3}\rangle\). Thus, the rapid decrease of $f_{E_0,1}$ leads to the drop of the parameter $A$ as $U$ increases from $0$.

Now let us briefly consider the region \(-0.15\,\text{eV}<U<0\,\text{eV}\). In this region,  we attribute the decrease of mini-gap $\Delta$ in Fig.\ref{Fig-2-dispersion-minigap}(c) to the rapid drop of the inverted gap $M$. In this region, it turns out that the changes of \(A\) and \(\xi_e\) are not significant ($\frac{A(U=-0.15)}{A(U=0)}=1.6$ and $\frac{\xi_e(U=-0.15)}{\xi_e(U=0)}=1.4$), while the inverted gap decreases greatly $\frac{M(U=-0.15)}{M(U=0)}=0.04$, leading to a reduction of mini-gap $\Delta$ ($\frac{\Delta(U=-0.15)}{\Delta(U=0)}=0.32$).

\textit{Conclusions and discussions-}
In summary, electric field dependences of the mini-gap and inverted gap are carefully studied for InAs/GaSb QWs based on the four band model and more sophisticated Kane model. Our results show opposite behaviors of the mini-gap and the inverted gap in certain parameter region, which provides a possible explanation of the puzzling observations in recent experiments of this system \cite{fanming_prl_2015,dulingjie2015}. We also notice recent debates about the origin of the quantized conductance plateau and the observation of edge transport in the trivial insulator side \cite{susanne_prb_2015}, as well as the identification of inverted or normal band gap \cite{fanming_prl_2015,fabrizio_arxiv_2015}. In particular, the behaviors of mini-gap $\Delta$ under an in-plane magnetic field are still not well understood \cite{fanming_prl_2015,fabrizio_arxiv_2015,dulingjie2015}. We hope our calculations can shed light on the underlying physics behind these controversial experimental observations.

\textit{Acknowledgement -}
We acknowledge the helpful discussion with R. R. Du for setting down this project. YZ is supported by National Basic Research Program of China (No.2014CB921201), NSFC (No.11374256) and the Fundamental Research Funds for the Central Universities in China. FCZ is supported by National Basic Research Program of China (No.2014CB921203) and NSFC (No.11274269). C.-X.L. acknowledges the support from Office of Naval Research (Grant No. N00014-15-1-2675).

\bibliography{reference}

\clearpage

\begin{widetext}

\begin{center}
\textbf{Supplementary Materials for $''$Electric control of inverted gap and hybridization gap in type II InAs/GaSb quantum wells$''$}
\end{center}

\section{Appendix I: Eight-band Kane model and Method}
  This QW system can be well described by the 8-band Kane model\cite{Kane_JPCS_1957} at $\Gamma$ point in the framework of $\mathbf{k}\cdot\mathbf{p}$ theory, and we choose the usual bulk basis\cite{kane_book12,winkler_book} as
  \begin{align}
     \vert1\rangle &= \vert\Gamma^6,1/2\rangle = \vert S\rangle\vert\uparrow\rangle \nonumber \\
     \vert2\rangle &= \vert\Gamma^6,-1/2\rangle = \vert S\rangle\vert\downarrow\rangle \nonumber \\
     \vert3\rangle &= \vert\Gamma^8,3/2\rangle = \frac{1}{\sqrt{2}}\vert X+iY\rangle\vert\uparrow\rangle \nonumber \\
     \vert4\rangle &= \vert\Gamma^8,1/2\rangle = \frac{1}{\sqrt{6}}\left(-2\vert Z\rangle\vert\uparrow\rangle + \vert X+iY\rangle\vert\downarrow\rangle \right) \nonumber \\
     \vert5\rangle &= \vert\Gamma^8,-1/2\rangle = -\frac{1}{\sqrt{6}}\left(2\vert Z\rangle\vert\downarrow\rangle + \vert X-iY\rangle\vert\uparrow\rangle \right) \nonumber \\
     \vert6\rangle &= \vert\Gamma^8,-3/2\rangle = -\frac{1}{\sqrt{2}}\vert X-iY\rangle\vert\downarrow\rangle \nonumber \\
     \vert7\rangle &= \vert\Gamma^7,1/2\rangle = \frac{1}{\sqrt{3}}\left(\vert Z\rangle\vert\uparrow\rangle + \vert X+iY\rangle\vert\downarrow\rangle\right) \nonumber \\
     \vert8\rangle &= \vert\Gamma^7,-1/2\rangle = \frac{1}{\sqrt{3}}\left(-\vert Z\rangle\vert\downarrow\rangle + \vert X-iY\rangle\vert\uparrow\rangle\right)
  \end{align}
  where we use the standard notation with $\vert\Gamma^6,\pm1/2\rangle$ representing an s-like conduction band, $\vert\Gamma^8,\pm1/2\rangle$ a p-like light hole band and $\vert\Gamma^8,\pm3/2\rangle$ a p-like heavy hole band in zinc blende crystal structure[\onlinecite{winkler_book}]. The SO split-off bands $\vert\Gamma^7,\pm1/2\rangle$ are far away in energy from the other bands, which will still be considered and kept in the calculation of electronic band structure; However, these two bands are not important at all and will be omitted to derive the effective 4-band BHZ model. Now, we would introduce our starting points, i.e., the 8-band Kane Hamiltonian $\mathcal{H}_{K}$, which is described anywhere, such as in Ref.\cite{novik_prb_2005,chaoxing_prl_2008,qingze_prl_2014}. In the above basis set $\vert\lambda\rangle (\lambda=1,2,\cdots,8)$, the hamiltonian $\mathcal{H}_{K}$ for two-dimensional system with [001] growth direction takes the following form:

    \begin{equation}\label{eq:ham_kane}
      \mathcal{H}_{\text{K}} = \left(
                          \begin{array}{cccccccc}
                            T & 0 & -\frac{1}{\sqrt{2}}Pk_{+} & \sqrt{\frac{2}{3}}Pk_z & \frac{1}{\sqrt{6}}Pk_{-} & 0 & -\frac{1}{\sqrt{3}}Pk_z & -\frac{1}{\sqrt{3}}Pk_{-} \\
                            0 & T & 0 & -\frac{1}{\sqrt{6}}Pk_{+} & \sqrt{\frac{2}{3}}Pk_z & \frac{1}{\sqrt{2}}Pk_{-} & -\frac{1}{\sqrt{3}}Pk_{+} & \frac{1}{\sqrt{3}}Pk_z \\
                            -\frac{1}{\sqrt{2}}k_{-}P & 0 & U+V & -\bar{S}_{-} & R & 0 & \frac{1}{\sqrt{2}}\bar{S}_{-} & -\sqrt{2}R \\
                            \sqrt{\frac{2}{3}}k_zP & -\frac{1}{\sqrt{6}}k_{-}P & -\bar{S}_{-}^{\dagger} & U-V & C & R & \sqrt{2}V & -\sqrt{\frac{3}{2}}\tilde{S}_{-} \\
                            \frac{1}{\sqrt{6}}k_{+}P & \sqrt{\frac{2}{3}}k_zP & R^{\dagger} & C^{\dagger} & U-V & \bar{S}_{+}^{\dagger} & -\sqrt{\frac{3}{2}}\tilde{S}_{+} & -\sqrt{2}V \\
                            0 & \frac{1}{\sqrt{2}}k_{+}P & 0 & R^{\dagger} & \bar{S}_{+} & U+V & \sqrt{2}R^{\dagger} & \frac{1}{\sqrt{2}}\bar{S}_{+} \\
                             -\frac{1}{\sqrt{3}}k_zP & -\frac{1}{\sqrt{3}}k_{-}P & \frac{1}{\sqrt{2}}\bar{S}_{-}^{\dagger} & \sqrt{2}V & -\sqrt{\frac{3}{2}}\tilde{S}_{+}^{\dagger} & \sqrt{2}R & U-\Delta & C \\
                            -\frac{1}{\sqrt{3}}k_{+}P & \frac{1}{\sqrt{3}}k_zP & -\sqrt{2}R^{\dagger} & -\sqrt{\frac{3}{2}}\tilde{S}_{-}^{\dagger} & -\sqrt{2}V & \frac{1}{\sqrt{2}}\bar{S}_{+}^{\dagger} & C^{\dagger} & U-\Delta \\
                          \end{array}
                        \right)
    \end{equation}
  where $k_{\parallel}^2 = k_x^2+k_y^2,\quad k_{\pm}=k_x\pm ik_y, \quad k_z=-i\partial/\partial z$, and the other elements are
  \begin{align}
     & T = E_c(z) + \frac{\hbar^2}{2m_0} \left\lbrack (2F+1)k_{\parallel}^2+k_z(2F+1)k_z \right\rbrack \nonumber \\
     & U = E_v(z) - \frac{\hbar^2}{2m_0}\left( \gamma_1 k_{\parallel}^2 + k_z\gamma_1 k_z \right)  \nonumber \\
     & V = -\frac{\hbar^2}{2m_0}\left( \gamma_2k_{\parallel}^2-2k_z\gamma_2k_z \right) \nonumber \\
     & R = -\frac{\hbar^2}{2m_0}\left( \sqrt{3}\mu k_{+}^2 -\sqrt{3}\bar{\gamma}k_{-}^2 \right) \nonumber \\
     & \bar{S}_{\pm} = -\frac{\hbar^2}{2m_0}\sqrt{3}k_{\pm}\left( \{\gamma_3,k_z\}+\lbrack \kappa,k_z \rbrack \right) \nonumber \\
     & \tilde{S}_{\pm} = -\frac{\hbar^2}{2m_0}\sqrt{3}k_{\pm}\left(  \{\gamma_3,k_z\} -\frac{1}{3}\left\lbrack \kappa,k_z \right\rbrack \right) \nonumber \\
     & C = \frac{\hbar^2}{m_0}k_{-}\lbrack \kappa,k_z \rbrack
  \end{align}
  $\lbrack A,B\rbrack=AB-BA$ is the usual commutator and $\{A,B\}=AB+BA$ is the usual anticommutator for the operators A and B; $P$ is the Kane momentum matrix element, and will be signed with a numerical value; $E_c(z)$ and $E_v(z)$ are the conduction and valence band edges, respectively; $\gamma_1,\gamma_2,\gamma_3,\kappa$, and $F$ represents the coupling to the remote bands, and the $\mu$ and $\bar{\gamma}$ are defined as $\mu=(\gamma_3-\gamma_2)/2$ and $\bar{\gamma}=(\gamma_3+\gamma_2)/2$; $\Delta$ is the spin-orbit splitting energy. The band structure parameters for InAs, GaSb and AlSb separately at $T=0 K$ are listed in Table \ref{tab:parameters_band}.

  \begin{table*}[!htbp]
    \caption{\label{tab:parameters_band} The band parameters for AlSb/InAs/GaSb/AlSb QWs in Kane model.}
    \begin{ruledtabular}
        \begin{tabular}{cccccccccc}
             & a($\text{\r{A}}$) & $E_c(eV)$ & $E_v(eV)$ & $\Delta(eV\cdot\text{\r{A}})$ & $P(eV\cdot\text{\r{A}})$ & $\gamma_1$ & $\gamma_2$ & $\gamma_3$ & $\kappa$  \\ \hline
            InAs & 6.0583 & -0.15  & -0.56 & 0.38  & 9.19 & 1.62 & -0.65 & 0.27 & -0.005 \\
            GaSb & 6.082  & 0.8128 & 0     & 0.752 & 9.23 & 2.61 & -0.56 & 0.67 & -0.23  \\
            AlSb & 6.133  & 1.94   & -0.38 & 0.75  & 8.43 & 1.46 & -0.33 & 0.41 & -0.23
        \end{tabular}
    \end{ruledtabular}
  \end{table*}

  Then, we would like to solve the 1D eigenvalues problem of Hamiltonian in Eq.(\ref{app-ham_full}) along z direction, since both $k_x$ and $k_y$ are good quantum number. Given $(k_x,k_y)$, we only need to solve this equation (\(k_z\to-i\partial_z\))
  \begin{align}\label{app-ham_full}
     \mathcal{H}_{\text{full}}(k_z)\vec{f}(z) = E \vec{f}(z)
  \end{align}
  where $\vec{f}(z)$ is the z direction envelope function vector $\left(f_1(z),f_2(z),\cdots,f_8(z)\right)^{T}$, the corresponded eigenvectors with eigenvalues $E^{\mu}$ can be expanded in the basis $\vert\lambda\rangle$ as
  \begin{align}
     \vert E^{\mu}\rangle = \sum_{\lambda=1}^{8} f_{\lambda}^{\mu}(z) \vert \lambda \rangle
  \end{align}

  Recall that $\vert\lambda\rangle$ are the eight basis for Kane model. And, we take plane-wave expansion to get the envelope functions
  \begin{align}
     f_{\lambda}^{\mu}(z) = \frac{1}{\sqrt{L}} \sum_{n} a_{n,\lambda}^{\mu} e^{ik_nz}
  \end{align}
  here $k_n=\frac{2\pi}{L}n, n=0,\pm1,\pm2,\cdots$. Therefore, we just need diagonalize a matrix to get the coefficients for envelope functions,
  \begin{align}\label{app-eq:eig_problem-matrix-ham}
     \sum_{m} \left\lbrack \mathcal{H}_{\text{full}}(k_z) \right\rbrack_{mn} \left\lbrack \vec{a}_{n}^{\mu} \right\rbrack = E^{\mu} \left\lbrack \vec{a}_{m}^{\mu} \right\rbrack
  \end{align}
  where we would like to take a cutoff for the total number of plane wave we picked in the diagonalization, $n\leq N_{\text{cut}}$. In practice, $N_{\text{cut}}=40$ is enough to get the low energy dispersion. The electronic properties of such systems depend strongly on the growth direction, our calculation only focuses on the [001] orientation in the following section.

\section{Appendix II: Critical width for InAs/GaSb Quantum wells}
  We focus on the critical quantum well thickness\cite{chaoxing_prl_2008} to find that the lowest subbands(InAs, denoted as $\vert E_0\rangle$) is crossed with the lowest subbands (GaSb, denoted as $\vert HH_0\rangle$), see Fig.\ref{Fig-4-appen-dc}(a)(b). Therefore, we will choose \(d_1=d_2=100 \text{ \r{A}}\) and \(d_3=250\text{ \r{A}}\) to numerical calculation in the main text, where we discuss the hybridized mini-gap \(\Delta\) as a function of the external electric potential \(U\).
 \begin{figure}[!htbp]
    \includegraphics[width=6in]{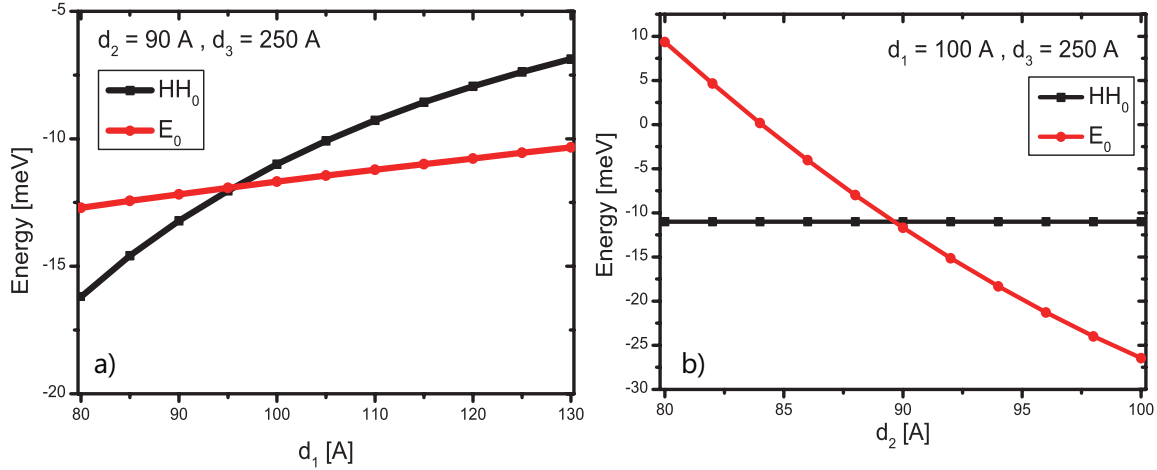}
    \caption{\label{Fig-4-appen-dc} a) Critical width of QWs for InAs by fixing the width of GaSb(10 nm) and AlSb(25 nm). b) Critical width of QWs for GaSb by fixing the width of InAs(9 nm) and AlSb(25 nm).}
\end{figure}

\section{Appendix III: Interface electric field}
  We would like to consider the effect of external electric field and interface electric field, the whole Hamiltonian reads
  \begin{align}\label{app-eq-ham_full}
     \mathcal{H}_{\text{full}} &= \mathcal{H}_{\text{K}} + \mathcal{V}_{\text{ext}} + \mathcal{V}_{\text{int}} \\
     \mathcal{V}_{\text{ext}}  &= \left\lbrack \left(\frac{U}{L}\right) \text{eV}\cdot\text{\r{A}}^{-1} \right\rbrack (z\,\text{\r{A}}) \\
     \mathcal{V}_{\text{int}}  &= \begin{cases}
                              \left\lbrack \left(\frac{U'}{2d_4}\right) \text{eV}\cdot\text{\r{A}}^{-1} \right\rbrack (z\,\text{\r{A}})\quad -d_4\leq z \leq d_4 \\
                              0 \quad\qquad\qquad\qquad\qquad\qquad \text{otherwise}
                           \end{cases}
  \end{align}
    \begin{figure}[!htbp]
    \includegraphics[width=6in]{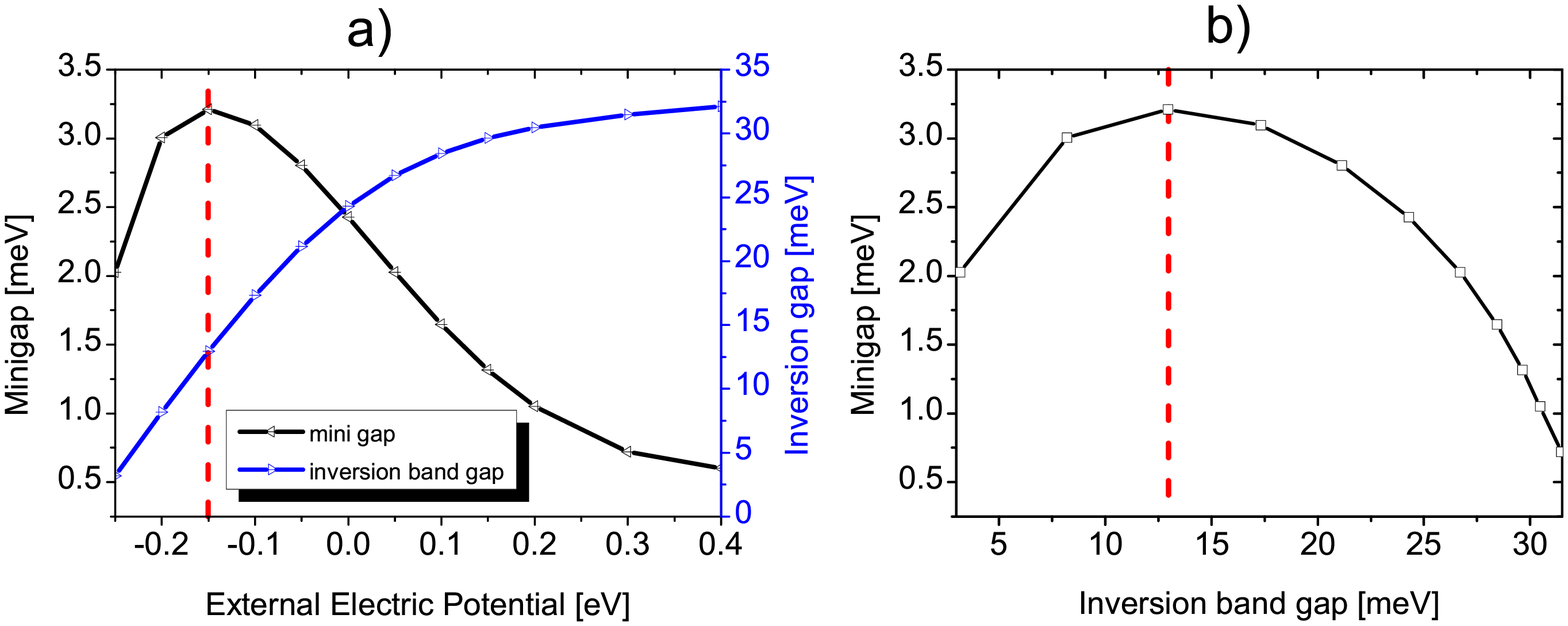}
    \caption{\label{Fig-5-appen-minigap} Consider the effect of both external and build-in electric field. It shows the relationship between the minigap/the inversion band gap with the external electric potential strength ($U$). a) The black line is for minigap, and the blue for inversion band gap. b) The relationship between minigap and inversion band gap.}
 \end{figure}
  where $L=d_1+d_2+2d_3$ is the whole width of the QWs; $U$ and $U'$ are the electric potential strength for external and interface, respectively; $d_4$ is the effective regime of the interface electric field, which is stem from the localized electron in the InAs layer and the localized hole in GaSb layer at the interface. The interface electric field is original from the structure asymmetry of AlSb/InAs/GaSb/AlSb QWs in Eq.(\ref{app-eq-ham_full}). Besides, we assume that $d_4<\text{min}\{d_1,d_2\}$. In practice, we chose \(d_4 = \frac{d_1+d_2}{4}\). Also, the interface electric field can be obtained through a self-consistent Poisson-Schrodinger calculation[\onlinecite{self-consis-1, self-consis-2, self-consis-3}]. However, we do not attempt to solve this self-consistent problem. Instead, we have phenomenologically assumed an effective electric potential occurring at the interface between InAs and GaSb.

  Next, we fix the interface electric field(internal): \(U'=0.05 \text{ eV}\) (strong enough), and we tune the external electric field(by tuning front gate and back gate), we get the relation between mini-gap \(\Delta\) and inversion band gap \(E_g\) with the electric field potential \(U\). Compare Fig 2 in main text (without interface electric field) and Fig.\ref{Fig-5-appen-minigap} below, we may find the physics is not changed, expect that the "critical" $U$ moves to negative from zero. The results for mini-gap and inversion band gap is similar to the results and explanation in the main text. Moreover, the external electric potential ($U$) will destroy the build-in interface electric potential ($U'$) by charge transfer. \\

\end{widetext}


\end{document}